%% file: TR.tex
\documentclass[a4paper,twocolumn,11pt]{article}
\usepackage[dvips]{color}
\usepackage[section]{algorithm}
\usepackage{algorithmic}
\usepackage{latexsym}
\usepackage{graphicx}
\usepackage{abstract}

\def\N{\mathrm{I\kern-.23emI\kern-.29em N}}
\newtheorem{theorem}{Theorem}
\newenvironment{proof}{{\par\noindent\bf Proof.}}%
	{\proofendbox\par\smallskip}
\newcommand{\proofendbox}{\penalty-1\mbox{}\hfill\lower.17ex\hbox{$\Box$}}

\title{Shortest Paths with Pairwise-Distinct Edge Labels:\\ 
{\Large Finding Biochemical Pathways in Metabolic Networks}}

\author{S\'andor Fekete\thanks{%
    Braunschweig University of Technology,
    Department of Computer Science,
    Algorithms Group,
    38106 Braunschweig, Germany.  
    \texttt{s.fekete@tu-bs.de, tom@kamphans.de, michaelstelzer@freenet.de}
}
\and	Tom Kamphans\footnotemark[1]
\and	Michael Stelzer\thanks{%
  Helmholtz Centre for Infection Research (HZI), 
  Systems Biology, 
  38124 Braunschweig, Germany
}
}

\date{}

\begin{document}
\maketitle
\begin{abstract}
\noindent
A problem studied in Systems Biology is how to
find shortest paths in metabolic networks.
Unfortunately, simple (i.e., graph theoretic) shortest paths do
not properly reflect biochemical facts. An approach to overcome
this issue is to use edge labels and search for paths with
distinct labels.

In this paper, we show that such biologically feasible shortest paths
are hard to compute. Moreover, we present solutions
to find such paths in networks in reasonable time.

\end{abstract}

\section{Introduction\label{intro}}


In the past decade, a lot of work has been done in the field of
bioreaction databases, which are a powerful tool for studying
biochemical processes (see, for example, Francke et al.~\cite{fst-rmnbi-05}).
Examples for such databases are the {\em Roche wall chart of
Biochemical Pathways}~\cite{e-bp-05}, the {\em BioCyc}
databases (e.g.~\cite{caddfgkkkklmpppszk-mdmpe-10}),
the {\em Kyoto Encyclopedia of Genes and Genomes} databases ({\em KEGG})~\cite{kgfth-kramn-10}
and the {\em BRaunschweig ENzyme
DAtabase BRENDA}~\cite{sgcsmrssts-beis2-11}.

\begin{figure}[t!]
\centering
\includegraphics{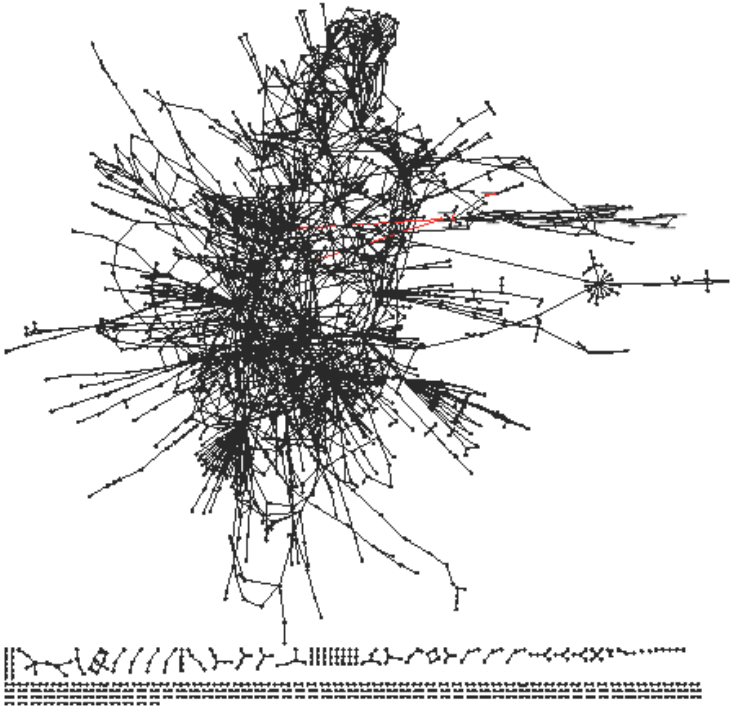}
\caption{Example of an organism-specific metabolic network ({\em
  H}. {\em sapiens}), reconstructed from a bioreaction
database~\cite{sszfk-ebdts-}.\label{network}}
\end{figure}

Biochemical reaction databases allow the reliable reconstruction of metabolic
networks, see Figure~\ref{network}, 
which---in turn---help to improve the annotation of
genome sequences to interpret high-throughput omics data, to
understand biological processes of pathogenesis or industrial
production at a system level, and to rationally control or design
biological systems 
(e.g., \cite{dhp-rveif-04,jks-esada-06,rs-olgpm-06,slrz-mpand-07}).

Such databases are subject to perpetual research and update: New
insights lead to extensions and corrections, resulting in more
accurate databases and more detailed views on biological aspects.
Under certain circumstances, database updates can be evaluated by
considering shortest paths in metabolic networks that have been
reconstructed from the 
database~\cite{mz-rmngd-03,r-btvsa-06,rsmhz-mn-08}.  
In the case of a metabolic network, a shortest path describes the
number of necessary reactions to convert one metabolite into 
another---provided that the network does not contain so-called currency
metabolites, which would introduce biologically infeasible shortcuts.
Ma and Zeng~\cite{mz-rmngd-03} developed such a database, and
demonstrated its usefulness by reconstruction of high-resolution
metabolic networks. In this database, Ma and
Zeng defined the reversibility of the reactions according to
literature data and biochemistry know\-ledge and introduced the concept
of reactant pairs by considering currency metabolites.
(The background for our work is a recent
upgrade of this database~\cite{sszfk-ebdts-}.)

Furthermore, shortest paths are of
particular interest in organism-specific metabolic networks, because
they can help to find alternative pathways (e.g.,~\cite{rjphss-mnaia-05}).
The tool {\em gapFiller} uses shortest paths to detect
gaps that occur in biochemical pathways (i.e., biomass that is
produced by an organism but not documented in the network due to a
lack of complete annotation) \cite{l-pc-10}. Additionally, 
shortest paths can be used to compute centrality indices~\cite{b-fabc-01}.

From a computer science point of view, shortest paths in a graph are
easy to compute using {\em Breadth-First Search} or {\em
Dijkstra's algorithm}; see, for example, Cormen et al.~\cite{clrs-ia-01}. 

However, simple (i.e., graph-theoretic) shortest paths do
not properly reflect biochemical facts: Biochemical pathways in
metabolic networks lead from one node (representing a
metabolite) to another one. The edges passed on this pathway 
represent biochemical reactions that convert a substrate to a product. 
Given a
reaction that leads from a substrate to two different products and
back--such as from 
{\em 2-dehydro-3-deoxy-D-galactonate 6-phosphate} (KEGG compound ID C01286) to 
{\em glyceraldehyde 3-P} (C00118) and {\em pyruvate} (C00022)---it is
possible to find a shortest path that leads from one product via the
substrate to the other one. One way to avoid this effect is to
store the reaction types as labels on the edges of the network, and
ensure that a shortest path passes no label twice \cite{r-btvsa-06,rsmhz-mn-08}.

Thus, this work considers shortest paths with pairwise-distinct edge 
labels for computing  {\em biologically
feasible} shortest paths (in metabolic networks that have been
reconstructed based on a database without currency metabolites and 
with reversibility information). 
In particular, 
we show in Section~\ref{spul-comp} 
that this kind of shortest paths is hard to compute.


\subsection*{Related Work}
Shortest paths have been considered in many settings and applications such
as robotics and traffic optimization.
For an overview see the survey articles by Mitchell \cite{m-spn-97,m-gspno-00}.
If the graph is unknown, common solutions are the
$A^*$- or $D^*$-algorithms~\cite{hnr-fbhdm-68,s-oeppp-94}.
Fleischer et al.~\cite{fkklt-colao-08} give a general framework for computing
short paths for settings where the environment as well as the 
location of the target is unknown.

For finding biologically feasible shortest paths, the approach by Faust
et al.~\cite{fch-mpura-09} is to take the chemical structure of
reactants into account to differentiate between side and main
compounds of a reaction.  Croes et al.~\cite{ccwh-mpirp-05} use
shortest paths in networks where a compound is assigned a weight equal
to its number of incident edges.
Finding pathways based on atomic transfers was presented by Boyer and
Viari~\cite{bv-airmp-03} and Heath et al.~\cite{hbk-fmpua-10}.  McShan
et al.~\cite{mrs-ppmph-03} use heuristic search methods for finding
metabolic pathways. Kaleta et al.~\cite{kfs-cwbls-09}
compute elementary flux patterns to detect pathways.

Independently from our work, Chakraborty et al.~\cite{cfmy-harc-09}
show that it is NP-hard to find the minimum number of labels such that any 
two vertices are connected by a {\em rainbow path} (i.e., a path with pairwise 
distinct edge labels).

There are many tools available for computing shortest paths and
analyzing biological networks in
various contexts. The tool {\em Pajek} \cite{bm-pplna-98} has its
roots in Social Sciences. The origin of {\em Cytoscape} 
(e.g.,~\cite{ccbsaklwicsbs-cseim-05}) 
is Systems Biology. The main focus of
both tools is on network visualization but both allow also some
analysis of networks.
Another program is the {\em Pathway Hunter Tool}
\cite{rasss-mpaws-05}.
All these tools allow for the computation of
shortest paths but none is---so far---able to compute
shortest paths with unique labels.
Software for computing such paths
is available from the authors~\cite{ks-sspul-08}.

\section{Background from Computer Science}\label{backInf}

In this section, we briefly review some basics in computer science
for readers with more background in biology than in computer science.

\subsection{Modeling Metabolic Networks}
Mathematically, a metabolic network is a graph, so let us briefly
review some terms from graph theory. A graph consists of a set of 
{\em vertices} or {\em nodes}. Relations between two nodes are modeled by
edges between nodes. If relations are unidirectional (i.e., the
edges are arrows), the graph is called {\em directed} graph.

In our case, the nodes represent the metabolites, the edges model
reactions converting one metabolite into another. Usually, reactions are
reversible; that is, the corresponding edges are undirected. However,
there are a few cases where the reverse direction is not feasible from
thermodynamic aspects. As directed graphs are easier to handle than
mixed graphs, we use only directed edges, and model reversible reactions
by two oppositely directed edges.

A {\em path} in a graph is a sequence of edges, where---except for the
first edge---every edge starts in that node where the preceding edge
ends. The {\em length} of a path is defined as the number of edges on
the path (if the edges have no weights).  Consequently, the shortest
path between two given nodes in a graph is the path having the
smallest number of edges.

\begin{table}[b!]
\hrulefill\\[-20pt]
\begin{tabbing}
\mbox\qquad\=\qquad\=\qquad\=\qquad\=\kill
Let $Q$ be a queue of vertices\\
insert start vertex into $Q$\\
while $Q$ is not empty do\\
\> $v$ := first vertex from $Q$\\
\> remove first vertex from $Q$\\
\> {\bf for all} vertices $v'$ adjacent to $v$ {\bf do}\\
\>\> {\bf if} $v'$ was not visited before {\bf then}\\
\>\>\> {\em v'.father} := $v$\\
\>\>\> mark $v'$ as visited\\
\>\>\> append $v'$ to $Q$\\
\>\>\> report shortest path to $v'$\\
\>\> {\bf end if}\\
\> {\bf end for}\\
{\bf end while}\\[-25pt]
\end{tabbing}
\hrulefill\\
Algorithm 1: Breadth-First Search (BFS)
\end{table}

\subsection{Shortest Paths in Unlabeled Networks}
Given a graph, $G=(V, E)$, it is quite simple to compute the shortest
path between two vertices of the graph or even from a given vertex to
every other one.  {\em Breadth-First Search} (BFS), see
Algorithm~1, is known to every first-grade student of computer
sciences. Even if the edges have different length, the problem is
still easy to compute using {\em Dijkstra's algorithm} 
(e.g., Cormen et al.~\cite{clrs-ia-01}).

Note that BFS creates a tree of all shortest paths by storing for
every vertex, $v$, its predecessor, {\em v.father}, on the shortest
path to the start vertex. Thus, a shortest path from a given vertex to
the start vertex can be found simply by following the {\em father}
pointers.

\subsection{Computational Complexity}\label{compcomp}
\begin{figure}[t!]
\centering
\includegraphics[height=8cm]{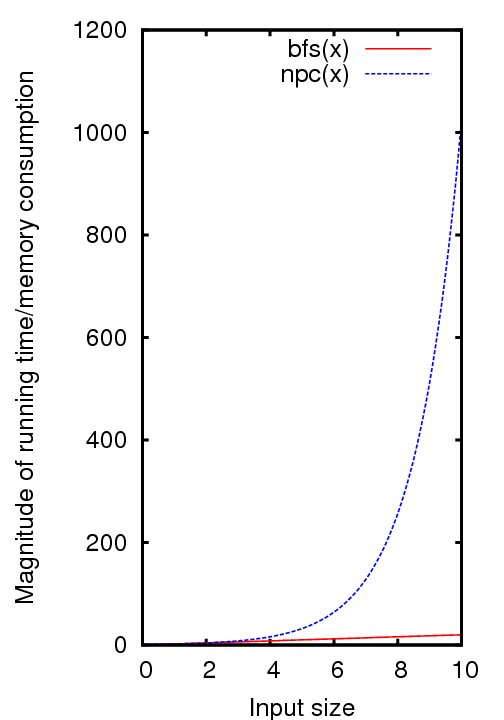}
\caption{Magnitude of resource request for an
NP-complete problem compared to BFS.\label{figs/exp-fig}}
\end{figure}

As we will see in the following section, things get much harder
in our case.
To express the hardness of a problem, it is very common in computer
science to estimate the order of growth in the resources (i.e.,
running time and memory requirements) needed by a program as a
function in the input size. In our case, the input size is the number
of vertices and edges in the graph. While the running time of BFS is
linear in the input size (that is, for example, if we double the input
size, the running time doubles also), we can show that the running
time is most likely to be exponential for shortest paths with unique labels;
that is, if we have a graph with $v$ vertices and $e$ edges, the
running time is in the order of $2^{v+e}$. More precisely, we can
show that our problem belongs to the class of {\em NP}-{\em complete}
problems~\cite{gj-cigtn-79}.  It is a widely held belief that there is
no sub-exponential solution for NP-complete problems. The impact of
this running time is shown in Figure~\ref{figs/exp-fig}.

NP-completeness is usually shown
using a common technique in computer science known
as proof by {\em reduction}. That is, we take a well-known hard
problem---in our case 3-SAT---and show that this problem would be easy
to solve if our problem {\em shortest path with unique labels problem} 
(SPUL), as defined in the following section, would
be easy to solve. This is done by describing how to translate an input
to \linebreak 3-SAT to SPUL such that a solution to SPUL yields a solution to
3-SAT. 

Given a set of {\em binary variables}, $x_1, \ldots, x_n$, and
a set of {\em clauses}, $C_1, \ldots, C_m$, consisting of three
literals (i.e., $C_i = L_{i1} \vee L_{i2} \vee L_{i3}$, where $L_{ik}$
denotes a negated ($\overline{x_k}$) or unnegated variable ($x_k$)),
the problem {\em 3-SAT} asks if there is a assignment
of $x_1, \ldots, x_n$ to 0 or 1 such that all clauses are 
fulfilled~\cite{gj-cigtn-79}.  


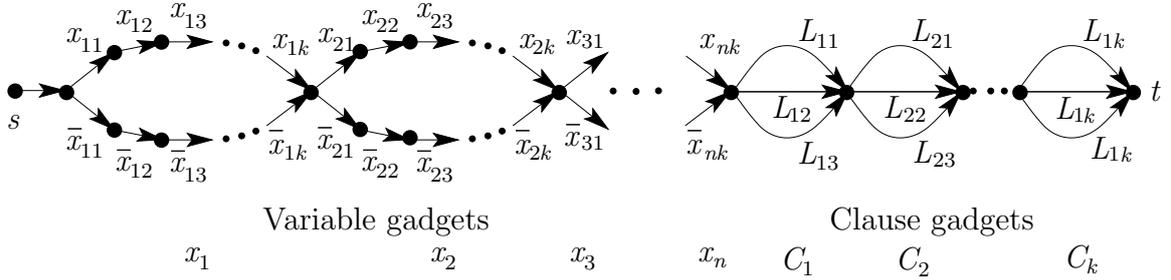
\begin{figure*}[t!]
\centerline{\input{figs/spul-np}}
\caption{Transforming a 3-SAT instance to an SPUL instance.
\label{figs/spul-np-fig}}
\end{figure*}

\section{Shortest Paths with Unique Labels}
\label{spnet} 

Let $G$ be a directed graph (in our
case a metabolic network) that consists of a set of vertices
(metabolites), $V$, and a set of edges (reactions), $E$.
Even if $G$ does not contain currency
metabolites, which would introduce biologically infeasible shortcuts,
a shortest path in $G$ may still be biologically infeasible, 
because it may use the same reaction
twice, as explained in Section~\ref{intro}.

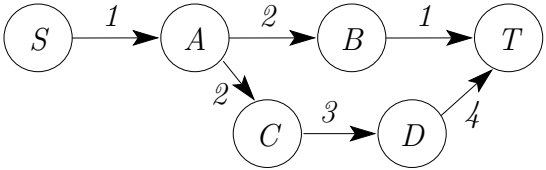
\begin{figure}[t]
\centerline{\input{figs/problem}}
\caption{The shortest path from $S$ to $T$ is
 $S\stackrel{1}{\longrightarrow} A
   \stackrel{2}{\longrightarrow} B
   \stackrel{1}{\longrightarrow} T$, but this path is
 infeasible, because it passes the label ``$1$'' twice. The shortest
 feasible path is \newline
 $S\stackrel{1}{\longrightarrow} A
   \stackrel{2}{\longrightarrow} C
   \stackrel{3}{\longrightarrow} D 
   \stackrel{4}{\longrightarrow} T$.
\label{figs/problem-fig}}
\end{figure}

Thus, we are interested only in shortest {\em feasible} paths; that
is, shortest paths from $s$ to $t$ with distinct reaction types. Such 
a path may be longer than the overall 
shortest path, see Figure~\ref{figs/problem-fig}. To store the
reaction types in the graph, we use {\em labels} for the edges.
Altogether,
a mathematical model for our problem is:

{\bf Problem} {\em Shortest Path with Unique Labels} (SPUL): Given a graph
$G=(V, E)$, a mapping $\ell: E\longrightarrow \N$ that
assigns a label to every edge, and two vertices, $s\in V$ and $t\in V$,
find a shortest path $P=(e_1, e_2, ..., e_k)$ starting
in $s$ and ending in $t$ with pairwise distinct edge labels; that is, for
$1\leq i < j \leq k: \ell(e_i) \neq \ell(e_j)$.  

\subsection{Computational Complexity}\label{spul-comp}
In an unlabeled network, we can simply apply BFS.
Things get considerably harder, if we add labels to the edges
and require that no label is passed twice on a path between two
vertices. Note that neither BFS nor Dijkstra's algorithm are
able to find the shortest feasible path shown in Figure~\ref{figs/problem-fig}.  

\begin{theorem}
Given a graph $G=(V, E)$ with a mapping
$\ell:E\longrightarrow \N$ that assigns a label to every edge, it is
NP-complete to determine if there is a path 
$P=(e_1, e_2, \ldots, e_k)$ that uses every label at most once; that is,
for $1\leq i< j \leq k: \ell(e_i) \neq \ell(e_j)$.
\end{theorem}
\begin{proof}
We show our theorem by reduction from 3-SAT.
A 3-SAT instance can be transformed to a SPUL instance as follows: For
every clause $C_i$ we use a clause gadget that consists of three
parallel edges labelled with $L_{ik}$ (i.e., with $x_{ik}$ or
$\overline{x_k}$ for an unnegated or negated variable,
respectively). The variable gadget for variable $x_j$ consists of two
parallel paths, one with all negated labels, one with all unnegated
labels. For the whole input, we start with a vertex, $s$, and add all
variable gadgets followed by all clause gadgets. The last vertex is
labeled $t$; see Figure~\ref{figs/spul-np-fig}. To find a path from $s$ to $t$, we
have to pass either the negated or the unnegated branch for every
variable. Thus, after passing the variable gadgets we have either all
negated or all unnegated variables left to pass the clause gadgets
without using a label twice. This is possible if and only if the given
formula is satisfiable.
\end{proof}

\begin{table}[t!]
\hrulefill\\[-20pt]
\begin{tabbing}
\mbox\quad\=\quad\=\quad\=\quad\=\quad\=\kill
Let $Q$ be a queue of edges\\
insert ``dummy edge'' to start vertex into $Q$\\
{\bf while} $Q$ is not empty {\bf do}\\
\> $e$ := first edge from $Q$\\
\> remove first edge from $Q$\\
\> {\bf for all} edges $e'$ adjacent to {\em e.target} {\bf do}\\
\>\> {\bf if} {\em $e'$.label} was not used on the shortest\\
\>\>\>\>  path from $s$ to $e$ {\bf then}\\
\>\>\> {\em $e'$.father} := $e$\\
\>\>\> append $e'$ to $Q$\\
\>\>\> {\bf if} {\em $e'$.target} was not visited before {\bf then}\\
\>\>\>\> report shortest path to {\em $e'$.target}\\
\>\>\> {\bf end if}\\
\>\> {\bf end if}\\
\> {\bf end for}\\
{\bf end while}\\[-25pt]
\end{tabbing}
\hrulefill\\
Algorithm 2: Shortest Path with Unique Labels
\end{table}

\subsection{Computing Shortest Path with Unique Labels}\label{computing}

\subsubsection{A Memory-Consuming Solution}\label{alg1}
\begin{figure}[t!]
\mbox{}\hspace{-1em}\includegraphics{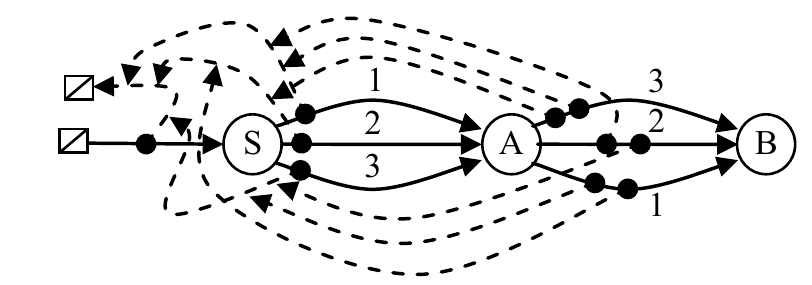}
\caption{An example for a shortest feasible path tree constructed by
  Algorithm~2. The network consists of three nodes (S, A, and B),
  three edges from S to A, and three edges from A to B. The dashed
  lines show the shortest feasible path tree.\label{ds}}
\end{figure}

We use a modified version of BFS to solve our problem, see
Algorithm~2. Similar to BFS, we store all paths found so far. But instead
of storing a shortest-path tree for the vertices as in BFS, we
construct a shortest-feasible-path tree on the edges of the graph, 
see Figure~\ref{ds}.
That is, we store every possible feasible path leading to a vertex in
the tree during the search. When the search reaches a vertex, $v$, via
an edge, $e'$, we can determine if the path to $v$ via $e'$ is
feasible (i.e., no label occurs twice). By the BFS-manner of this
algorithm, the first feasible path found to a vertex is also the
shortest feasible path. The drawback is that the algorithm is quite
memory consuming, because it stores all feasible paths to all
vertices.

\subsubsection{Balancing Time and Memory Requirements}
To save memory, we tried a different solution by exploiting the fact
that a metabolic network has many parallel edges. Thus, our search
does not have to explore all edges incident to a vertex, but only
those edges that lead to different vertices. Instead of storing one
label per edge on a shortest path, we store a set of labels. This
significantly decreases the number of shortest paths that we have to
store. The drawback is that we have to find a feasible combination of
labels when the search progresses (i.e., when we want to add a new
edge to a path). This can be solved by a simple backtracking; that is,
we successively test combinations of labels until we either find a
feasible combination or no more combination is possible. The idea is
that in most cases this backtracking may not require much time,
because a feasible combination is found quickly. Only if there is no
feasible combination, we have to test all of them. Clearly, this
heavily depends on the structure of the input network.

\subsubsection{Preprocessing}
Before we start our algorithm, we perform a simple BFS to determine,
which vertices can be reached at all (i.e., there is a feasible or
infeasible path). We store these vertices to be able to abort the
search as soon as feasible paths to all of them have been found. In a
second stage of the preprocessing, we perform a simple BFS again, this
time checking if the found path is feasible and reporting
feasible paths.

\def\TabShrink{}
\def\TabS{$\!\!\!\!\!$}

\begin{table*}
\caption{Examples of running time and found paths starting in vertex 1 (XEON CPU 3.0 GHz, 16 GB memory).\label{runtimes}\newline}

\noindent
\begin{tabular}{l|*{9}{c}}
& &  & \TabS Vertices\TabS 
& \multicolumn{3}{c}{Paths found} 
& \multicolumn{3}{c}{Running time in min}\TabShrink \\
Network & Vertices\TabShrink & Edges\TabShrink & reachable\TabShrink 
& Alg.\,A\TabS & Alg.\,B\TabShrink & B with\TabShrink 
& Alg.\,A\TabS & Alg.\,B\TabShrink & B with\TabShrink 
\\ 
& in $G$ & in $G$ & from start &&&preproc. &&&preproc.\\ \hline
{\em A.~niger} & 2547 & 7818 & 1488 & 1283 & 1369 & 1382 & 1.2 & 23.4 & 23.6 \\ 
{\em E.~coli} & 1895 & 5525 & 1111 & 911 & 1012 & 1021 & 1.3 & 35.6 & 35.9 \\
{\em H.~sapiens}\TabShrink & 2474 & 7873 & 1614 & 1347 & 1507 & 1526 & 1.2 & 20.2 & 20.5 \\
\end{tabular}
\end{table*}

\def\TabShr{$\!\!$}

\begin{table*}
\caption{Comparison on the number of paths found in several organism-specific metabolic networks\label{smaller}: Total number of shortest paths (SP),  
number of correct/infeasible shortest paths found by BFS,
number of shortest paths with unique labels (SPUL).
\newline
}

\noindent
\begin{tabular}{l|*{8}{c}}
 \TabShr & uur$^{\rm -cm}$ \TabShr & \TabShr uur$^{\rm +cm}$ \TabShr & \TabShr mpn$^{\rm -cm}$ \TabShr & \TabShr mpn$^{\rm +cm}$ \TabShr & \TabShr bbu$^{\rm -cm}$ \TabShr & \TabShr bbu$^{\rm +cm}$ \TabShr & \TabShr mge$^{\rm -cm}$ \TabShr & \TabShr mge$^{\rm +cm}$ \\ \hline
SP & 2372 & 17038 & 2191 & 6652 & 6357 & 39552 & 6793 & 36246\\ 
Correct SP & 1259 & 9302 & 1288 & 3929 & 2470 & 16864 & 2868 & 18460 \\
Infeasible SP & 1113 & 7736 & 903 & 2723 & 3887 & 22688 & 3925 & 17786\\
SPUL & 1308 & 13306 & 1601 & 4577 & 2513 & 22809 & 3061 & 22281\\[5pt]
\end{tabular}

{\footnotesize
uur = {\em U.\,urealyticum}, mpn = {\em M.\,pneumoniae}, bbu = {\em B.\,burgdorferi}, mge = {\em M.\,genitalium},\newline
+cm/-cm: with/without currency metabolites}
\end{table*}

\subsubsection{Comparison}
Table~\ref{runtimes} shows results for large databases. There was not sufficient
memory to compute all paths. Thus, we compare the number of paths
found until the program was stopped because there was no memory
left. It turned out that the memory consuming solution (Alg.~A) is
much faster than our second approach, but finds less paths. The
preprocessing with BFS further improves the number of found
paths. Smaller organisms are compared in Table~\ref{smaller}:
We compared the number of (graph theoretically)
shortest paths (SP) to the number of shortest paths with unique labels
(SPUL). Furthermore, we listed the 
number of correct shortest paths and the 
number of biologically infeasible
shortest paths
(i.e., paths such as 
S$\longrightarrow$ A$\longrightarrow$ B $\longrightarrow$ T in
Fig.~\ref{figs/problem-fig})
that were found using BFS.  

\section{Conclusion}
We showed that the problem of finding paths pairwise-distinct edge labels
in (directed) graphs is NP-hard. A straightforward modification of BFS to
this problem results in a memory-consuming solution that allows for computing only small
instances. Balancing time- and memory requirements, we are able to 
deal with larger instances.

\section*{Acknowledgements}
This work was financially supported by the {\em German Research
  Foundation} ({\em DFG}), project {\em SFB 578} and by {\em Federal
  Ministry of Education and Research} ({\em BMBF}), project {\em
  InterGenomics}.  Tom Kamphans was supported by 7th Framework
Programme contract 215270 (FRONTS).  

\bibliographystyle{abbrv}
{\small 
\bibliography{lit}
}

\end{document}

%% file: figs/spul-np.tex
\begin{picture}(0,0)%
\includegraphics{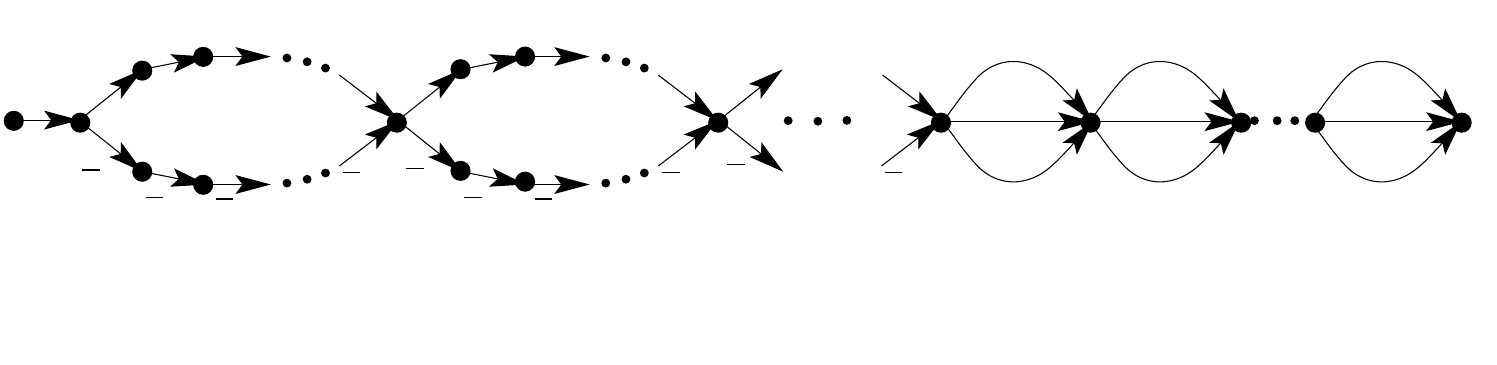}%
\end{picture}%
\setlength{\unitlength}{2693sp}%
\begingroup\makeatletter\ifx\SetFigFont\undefined%
\gdef\SetFigFont#1#2#3#4#5{%
  \reset@font\fontsize{#1}{#2pt}%
  \fontfamily{#3}\fontseries{#4}\fontshape{#5}%
  \selectfont}%
\fi\endgroup%
\begin{picture}(10516,2557)(345,-3128)
\put(8668,-3113){\makebox(0,0)[lb]{\smash{{\SetFigFont{8}{9.6}{\rmdefault}{\mddefault}{\updefault}{\color[rgb]{0,0,0}2}%
}}}}
\put(10198,-3082){\makebox(0,0)[lb]{\smash{{\SetFigFont{8}{9.6}{\rmdefault}{\mddefault}{\itdefault}{\color[rgb]{0,0,0}k}%
}}}}
\put(9992,-3034){\makebox(0,0)[lb]{\smash{{\SetFigFont{12}{14.4}{\rmdefault}{\mddefault}{\itdefault}{\color[rgb]{0,0,0}C}%
}}}}
\put(10770,-1505){\makebox(0,0)[lb]{\smash{{\SetFigFont{12}{14.4}{\rmdefault}{\mddefault}{\itdefault}{\color[rgb]{0,0,0}t}%
}}}}
\put(8794,-2685){\makebox(0,0)[b]{\smash{{\SetFigFont{12}{14.4}{\rmdefault}{\mddefault}{\updefault}{\color[rgb]{0,0,0}Clause gadgets}%
}}}}
\put(6796,-3031){\makebox(0,0)[lb]{\smash{{\SetFigFont{8}{9.6}{\rmdefault}{\mddefault}{\itdefault}{\color[rgb]{0,0,0}n}%
}}}}
\put(6661,-2980){\makebox(0,0)[lb]{\smash{{\SetFigFont{12}{14.4}{\rmdefault}{\mddefault}{\itdefault}{\color[rgb]{0,0,0}x}%
}}}}
\put(5491,-2980){\makebox(0,0)[lb]{\smash{{\SetFigFont{12}{14.4}{\rmdefault}{\mddefault}{\itdefault}{\color[rgb]{0,0,0}x}%
}}}}
\put(5626,-3054){\makebox(0,0)[lb]{\smash{{\SetFigFont{8}{9.6}{\rmdefault}{\mddefault}{\updefault}{\color[rgb]{0,0,0}3}%
}}}}
\put(5601,-1953){\makebox(0,0)[lb]{\smash{{\SetFigFont{8}{9.6}{\rmdefault}{\mddefault}{\updefault}{\color[rgb]{0,0,0}31}%
}}}}
\put(5601,-1016){\makebox(0,0)[lb]{\smash{{\SetFigFont{8}{9.6}{\rmdefault}{\mddefault}{\updefault}{\color[rgb]{0,0,0}31}%
}}}}
\put(5251,-1069){\makebox(0,0)[lb]{\smash{{\SetFigFont{8}{9.6}{\rmdefault}{\mddefault}{\itdefault}{\color[rgb]{0,0,0}k}%
}}}}
\put(5220,-2041){\makebox(0,0)[lb]{\smash{{\SetFigFont{8}{9.6}{\rmdefault}{\mddefault}{\itdefault}{\color[rgb]{0,0,0}k}%
}}}}
\put(1846,-736){\makebox(0,0)[lb]{\smash{{\SetFigFont{12}{14.4}{\rmdefault}{\mddefault}{\itdefault}{\color[rgb]{0,0,0}x}%
}}}}
\put(1351,-798){\makebox(0,0)[lb]{\smash{{\SetFigFont{12}{14.4}{\rmdefault}{\mddefault}{\itdefault}{\color[rgb]{0,0,0}x}%
}}}}
\put(901,-961){\makebox(0,0)[lb]{\smash{{\SetFigFont{12}{14.4}{\rmdefault}{\mddefault}{\itdefault}{\color[rgb]{0,0,0}x}%
}}}}
\put(901,-1950){\makebox(0,0)[lb]{\smash{{\SetFigFont{12}{14.4}{\rmdefault}{\mddefault}{\itdefault}{\color[rgb]{0,0,0}x}%
}}}}
\put(1351,-2139){\makebox(0,0)[lb]{\smash{{\SetFigFont{12}{14.4}{\rmdefault}{\mddefault}{\itdefault}{\color[rgb]{0,0,0}x}%
}}}}
\put(1846,-2155){\makebox(0,0)[lb]{\smash{{\SetFigFont{12}{14.4}{\rmdefault}{\mddefault}{\itdefault}{\color[rgb]{0,0,0}x}%
}}}}
\put(3188,-1945){\makebox(0,0)[lb]{\smash{{\SetFigFont{12}{14.4}{\rmdefault}{\mddefault}{\itdefault}{\color[rgb]{0,0,0}x}%
}}}}
\put(3594,-2139){\makebox(0,0)[lb]{\smash{{\SetFigFont{12}{14.4}{\rmdefault}{\mddefault}{\itdefault}{\color[rgb]{0,0,0}x}%
}}}}
\put(4089,-2155){\makebox(0,0)[lb]{\smash{{\SetFigFont{12}{14.4}{\rmdefault}{\mddefault}{\itdefault}{\color[rgb]{0,0,0}x}%
}}}}
\put(4991,-1986){\makebox(0,0)[lb]{\smash{{\SetFigFont{12}{14.4}{\rmdefault}{\mddefault}{\itdefault}{\color[rgb]{0,0,0}x}%
}}}}
\put(5026,-1019){\makebox(0,0)[lb]{\smash{{\SetFigFont{12}{14.4}{\rmdefault}{\mddefault}{\itdefault}{\color[rgb]{0,0,0}x}%
}}}}
\put(4089,-736){\makebox(0,0)[lb]{\smash{{\SetFigFont{12}{14.4}{\rmdefault}{\mddefault}{\itdefault}{\color[rgb]{0,0,0}x}%
}}}}
\put(3594,-798){\makebox(0,0)[lb]{\smash{{\SetFigFont{12}{14.4}{\rmdefault}{\mddefault}{\itdefault}{\color[rgb]{0,0,0}x}%
}}}}
\put(3193,-961){\makebox(0,0)[lb]{\smash{{\SetFigFont{12}{14.4}{\rmdefault}{\mddefault}{\itdefault}{\color[rgb]{0,0,0}x}%
}}}}
\put(5153,-1070){\makebox(0,0)[lb]{\smash{{\SetFigFont{8}{9.6}{\rmdefault}{\mddefault}{\updefault}{\color[rgb]{0,0,0}2}%
}}}}
\put(5126,-2041){\makebox(0,0)[lb]{\smash{{\SetFigFont{8}{9.6}{\rmdefault}{\mddefault}{\updefault}{\color[rgb]{0,0,0}2}%
}}}}
\put(3729,-2211){\makebox(0,0)[lb]{\smash{{\SetFigFont{8}{9.6}{\rmdefault}{\mddefault}{\updefault}{\color[rgb]{0,0,0}22}%
}}}}
\put(4224,-2245){\makebox(0,0)[lb]{\smash{{\SetFigFont{8}{9.6}{\rmdefault}{\mddefault}{\updefault}{\color[rgb]{0,0,0}23}%
}}}}
\put(3323,-2023){\makebox(0,0)[lb]{\smash{{\SetFigFont{8}{9.6}{\rmdefault}{\mddefault}{\updefault}{\color[rgb]{0,0,0}21}%
}}}}
\put(3333,-1046){\makebox(0,0)[lb]{\smash{{\SetFigFont{8}{9.6}{\rmdefault}{\mddefault}{\updefault}{\color[rgb]{0,0,0}21}%
}}}}
\put(3729,-871){\makebox(0,0)[lb]{\smash{{\SetFigFont{8}{9.6}{\rmdefault}{\mddefault}{\updefault}{\color[rgb]{0,0,0}22}%
}}}}
\put(4224,-830){\makebox(0,0)[lb]{\smash{{\SetFigFont{8}{9.6}{\rmdefault}{\mddefault}{\updefault}{\color[rgb]{0,0,0}23}%
}}}}
\put(1981,-830){\makebox(0,0)[lb]{\smash{{\SetFigFont{8}{9.6}{\rmdefault}{\mddefault}{\updefault}{\color[rgb]{0,0,0}13}%
}}}}
\put(1486,-871){\makebox(0,0)[lb]{\smash{{\SetFigFont{8}{9.6}{\rmdefault}{\mddefault}{\updefault}{\color[rgb]{0,0,0}12}%
}}}}
\put(1036,-1051){\makebox(0,0)[lb]{\smash{{\SetFigFont{8}{9.6}{\rmdefault}{\mddefault}{\updefault}{\color[rgb]{0,0,0}11}%
}}}}
\put(1036,-2023){\makebox(0,0)[lb]{\smash{{\SetFigFont{8}{9.6}{\rmdefault}{\mddefault}{\updefault}{\color[rgb]{0,0,0}11}%
}}}}
\put(1486,-2211){\makebox(0,0)[lb]{\smash{{\SetFigFont{8}{9.6}{\rmdefault}{\mddefault}{\updefault}{\color[rgb]{0,0,0}12}%
}}}}
\put(1981,-2245){\makebox(0,0)[lb]{\smash{{\SetFigFont{8}{9.6}{\rmdefault}{\mddefault}{\updefault}{\color[rgb]{0,0,0}13}%
}}}}
\put(1977,-2980){\makebox(0,0)[lb]{\smash{{\SetFigFont{12}{14.4}{\rmdefault}{\mddefault}{\itdefault}{\color[rgb]{0,0,0}x}%
}}}}
\put(4220,-2980){\makebox(0,0)[lb]{\smash{{\SetFigFont{12}{14.4}{\rmdefault}{\mddefault}{\itdefault}{\color[rgb]{0,0,0}x}%
}}}}
\put(4357,-3054){\makebox(0,0)[lb]{\smash{{\SetFigFont{8}{9.6}{\rmdefault}{\mddefault}{\updefault}{\color[rgb]{0,0,0}2}%
}}}}
\put(2114,-3054){\makebox(0,0)[lb]{\smash{{\SetFigFont{8}{9.6}{\rmdefault}{\mddefault}{\updefault}{\color[rgb]{0,0,0}1}%
}}}}
\put(3731,-2685){\makebox(0,0)[b]{\smash{{\SetFigFont{12}{14.4}{\rmdefault}{\mddefault}{\updefault}{\color[rgb]{0,0,0}Variable gadgets}%
}}}}
\put(360,-1758){\makebox(0,0)[lb]{\smash{{\SetFigFont{12}{14.4}{\rmdefault}{\mddefault}{\itdefault}{\color[rgb]{0,0,0}s}%
}}}}
\put(2990,-2045){\makebox(0,0)[lb]{\smash{{\SetFigFont{8}{9.6}{\rmdefault}{\mddefault}{\itdefault}{\color[rgb]{0,0,0}k}%
}}}}
\put(3009,-1082){\makebox(0,0)[lb]{\smash{{\SetFigFont{8}{9.6}{\rmdefault}{\mddefault}{\itdefault}{\color[rgb]{0,0,0}k}%
}}}}
\put(2779,-1019){\makebox(0,0)[lb]{\smash{{\SetFigFont{12}{14.4}{\rmdefault}{\mddefault}{\itdefault}{\color[rgb]{0,0,0}x}%
}}}}
\put(2743,-1990){\makebox(0,0)[lb]{\smash{{\SetFigFont{12}{14.4}{\rmdefault}{\mddefault}{\itdefault}{\color[rgb]{0,0,0}x}%
}}}}
\put(2892,-2041){\makebox(0,0)[lb]{\smash{{\SetFigFont{8}{9.6}{\rmdefault}{\mddefault}{\updefault}{\color[rgb]{0,0,0}1}%
}}}}
\put(2915,-1083){\makebox(0,0)[lb]{\smash{{\SetFigFont{8}{9.6}{\rmdefault}{\mddefault}{\updefault}{\color[rgb]{0,0,0}1}%
}}}}
\put(5438,-1917){\makebox(0,0)[lb]{\smash{{\SetFigFont{12}{14.4}{\rmdefault}{\mddefault}{\itdefault}{\color[rgb]{0,0,0}x}%
}}}}
\put(5454,-985){\makebox(0,0)[lb]{\smash{{\SetFigFont{12}{14.4}{\rmdefault}{\mddefault}{\itdefault}{\color[rgb]{0,0,0}x}%
}}}}
\put(6715,-2041){\makebox(0,0)[lb]{\smash{{\SetFigFont{8}{9.6}{\rmdefault}{\mddefault}{\itdefault}{\color[rgb]{0,0,0}n}%
}}}}
\put(6827,-2041){\makebox(0,0)[lb]{\smash{{\SetFigFont{8}{9.6}{\rmdefault}{\mddefault}{\itdefault}{\color[rgb]{0,0,0}k}%
}}}}
\put(6826,-1070){\makebox(0,0)[lb]{\smash{{\SetFigFont{8}{9.6}{\rmdefault}{\mddefault}{\itdefault}{\color[rgb]{0,0,0}n}%
}}}}
\put(6938,-1069){\makebox(0,0)[lb]{\smash{{\SetFigFont{8}{9.6}{\rmdefault}{\mddefault}{\itdefault}{\color[rgb]{0,0,0}k}%
}}}}
\put(6668,-1019){\makebox(0,0)[lb]{\smash{{\SetFigFont{12}{14.4}{\rmdefault}{\mddefault}{\itdefault}{\color[rgb]{0,0,0}x}%
}}}}
\put(6553,-1986){\makebox(0,0)[lb]{\smash{{\SetFigFont{12}{14.4}{\rmdefault}{\mddefault}{\itdefault}{\color[rgb]{0,0,0}x}%
}}}}
\put(8598,-954){\makebox(0,0)[lb]{\smash{{\SetFigFont{12}{14.4}{\rmdefault}{\mddefault}{\itdefault}{\color[rgb]{0,0,0}L}%
}}}}
\put(10157,-954){\makebox(0,0)[lb]{\smash{{\SetFigFont{12}{14.4}{\rmdefault}{\mddefault}{\itdefault}{\color[rgb]{0,0,0}L}%
}}}}
\put(7740,-1018){\makebox(0,0)[lb]{\smash{{\SetFigFont{8}{9.6}{\rmdefault}{\mddefault}{\updefault}{\color[rgb]{0,0,0}11}%
}}}}
\put(10344,-1018){\makebox(0,0)[lb]{\smash{{\SetFigFont{8}{9.6}{\rmdefault}{\mddefault}{\updefault}{\color[rgb]{0,0,0}1}%
}}}}
\put(8785,-1018){\makebox(0,0)[lb]{\smash{{\SetFigFont{8}{9.6}{\rmdefault}{\mddefault}{\updefault}{\color[rgb]{0,0,0}21}%
}}}}
\put(10441,-1015){\makebox(0,0)[lb]{\smash{{\SetFigFont{8}{9.6}{\rmdefault}{\mddefault}{\itdefault}{\color[rgb]{0,0,0}k}%
}}}}
\put(7566,-954){\makebox(0,0)[lb]{\smash{{\SetFigFont{12}{14.4}{\rmdefault}{\mddefault}{\itdefault}{\color[rgb]{0,0,0}L}%
}}}}
\put(7331,-1657){\makebox(0,0)[lb]{\smash{{\SetFigFont{12}{14.4}{\rmdefault}{\mddefault}{\itdefault}{\color[rgb]{0,0,0}L}%
}}}}
\put(8349,-1656){\makebox(0,0)[lb]{\smash{{\SetFigFont{12}{14.4}{\rmdefault}{\mddefault}{\itdefault}{\color[rgb]{0,0,0}L}%
}}}}
\put(9915,-1663){\makebox(0,0)[lb]{\smash{{\SetFigFont{12}{14.4}{\rmdefault}{\mddefault}{\itdefault}{\color[rgb]{0,0,0}L}%
}}}}
\put(7489,-1698){\makebox(0,0)[lb]{\smash{{\SetFigFont{8}{9.6}{\rmdefault}{\mddefault}{\updefault}{\color[rgb]{0,0,0}12}%
}}}}
\put(8534,-1691){\makebox(0,0)[lb]{\smash{{\SetFigFont{8}{9.6}{\rmdefault}{\mddefault}{\updefault}{\color[rgb]{0,0,0}22}%
}}}}
\put(8804,-2110){\makebox(0,0)[lb]{\smash{{\SetFigFont{8}{9.6}{\rmdefault}{\mddefault}{\updefault}{\color[rgb]{0,0,0}23}%
}}}}
\put(8613,-2073){\makebox(0,0)[lb]{\smash{{\SetFigFont{12}{14.4}{\rmdefault}{\mddefault}{\itdefault}{\color[rgb]{0,0,0}L}%
}}}}
\put(7567,-2081){\makebox(0,0)[lb]{\smash{{\SetFigFont{12}{14.4}{\rmdefault}{\mddefault}{\itdefault}{\color[rgb]{0,0,0}L}%
}}}}
\put(7738,-2136){\makebox(0,0)[lb]{\smash{{\SetFigFont{8}{9.6}{\rmdefault}{\mddefault}{\updefault}{\color[rgb]{0,0,0}13}%
}}}}
\put(10508,-2061){\makebox(0,0)[lb]{\smash{{\SetFigFont{8}{9.6}{\rmdefault}{\mddefault}{\itdefault}{\color[rgb]{0,0,0}k}%
}}}}
\put(10402,-2056){\makebox(0,0)[lb]{\smash{{\SetFigFont{8}{9.6}{\rmdefault}{\mddefault}{\updefault}{\color[rgb]{0,0,0}1}%
}}}}
\put(10232,-2034){\makebox(0,0)[lb]{\smash{{\SetFigFont{12}{14.4}{\rmdefault}{\mddefault}{\itdefault}{\color[rgb]{0,0,0}L}%
}}}}
\put(10081,-1697){\makebox(0,0)[lb]{\smash{{\SetFigFont{8}{9.6}{\rmdefault}{\mddefault}{\updefault}{\color[rgb]{0,0,0}1}%
}}}}
\put(10169,-1702){\makebox(0,0)[lb]{\smash{{\SetFigFont{8}{9.6}{\rmdefault}{\mddefault}{\itdefault}{\color[rgb]{0,0,0}k}%
}}}}
\put(7411,-3047){\makebox(0,0)[lb]{\smash{{\SetFigFont{12}{14.4}{\rmdefault}{\mddefault}{\itdefault}{\color[rgb]{0,0,0}C}%
}}}}
\put(7611,-3107){\makebox(0,0)[lb]{\smash{{\SetFigFont{8}{9.6}{\rmdefault}{\mddefault}{\updefault}{\color[rgb]{0,0,0}1}%
}}}}
\put(8458,-3047){\makebox(0,0)[lb]{\smash{{\SetFigFont{12}{14.4}{\rmdefault}{\mddefault}{\itdefault}{\color[rgb]{0,0,0}C}%
}}}}
\end{picture}%

%% file: figs/problem.tex
\begin{picture}(0,0)%
\includegraphics{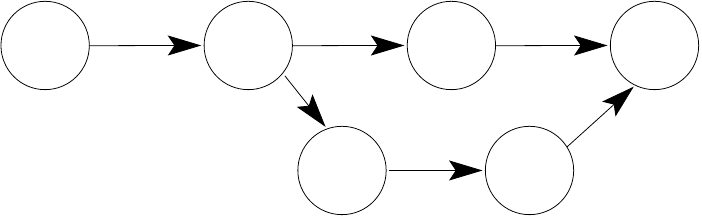}%
\end{picture}%
\setlength{\unitlength}{1973sp}%
\begingroup\makeatletter\ifx\SetFigFont\undefined%
\gdef\SetFigFont#1#2#3#4#5{%
  \reset@font\fontsize{#1}{#2pt}%
  \fontfamily{#3}\fontseries{#4}\fontshape{#5}%
  \selectfont}%
\fi\endgroup%
\begin{picture}(6714,2066)(1369,-3192)
\put(5401,-2611){\makebox(0,0)[b]{\smash{{\SetFigFont{12}{14.4}{\rmdefault}{\mddefault}{\itdefault}{\color[rgb]{0,0,0}3}%
}}}}
\put(1801,-1711){\makebox(0,0)[b]{\smash{{\SetFigFont{12}{14.4}{\rmdefault}{\mddefault}{\itdefault}{\color[rgb]{0,0,0}S}%
}}}}
\put(3751,-1711){\makebox(0,0)[b]{\smash{{\SetFigFont{12}{14.4}{\rmdefault}{\mddefault}{\itdefault}{\color[rgb]{0,0,0}A}%
}}}}
\put(5701,-1711){\makebox(0,0)[b]{\smash{{\SetFigFont{12}{14.4}{\rmdefault}{\mddefault}{\itdefault}{\color[rgb]{0,0,0}B}%
}}}}
\put(7651,-1711){\makebox(0,0)[b]{\smash{{\SetFigFont{12}{14.4}{\rmdefault}{\mddefault}{\itdefault}{\color[rgb]{0,0,0}T}%
}}}}
\put(4651,-2911){\makebox(0,0)[b]{\smash{{\SetFigFont{12}{14.4}{\rmdefault}{\mddefault}{\itdefault}{\color[rgb]{0,0,0}C}%
}}}}
\put(6451,-2911){\makebox(0,0)[b]{\smash{{\SetFigFont{12}{14.4}{\rmdefault}{\mddefault}{\itdefault}{\color[rgb]{0,0,0}D}%
}}}}
\put(2701,-1411){\makebox(0,0)[b]{\smash{{\SetFigFont{12}{14.4}{\rmdefault}{\mddefault}{\itdefault}{\color[rgb]{0,0,0}1}%
}}}}
\put(4651,-1411){\makebox(0,0)[b]{\smash{{\SetFigFont{12}{14.4}{\rmdefault}{\mddefault}{\itdefault}{\color[rgb]{0,0,0}2}%
}}}}
\put(6601,-1411){\makebox(0,0)[b]{\smash{{\SetFigFont{12}{14.4}{\rmdefault}{\mddefault}{\itdefault}{\color[rgb]{0,0,0}1}%
}}}}
\put(7201,-2611){\makebox(0,0)[b]{\smash{{\SetFigFont{12}{14.4}{\rmdefault}{\mddefault}{\itdefault}{\color[rgb]{0,0,0}4}%
}}}}
\put(4051,-2386){\makebox(0,0)[b]{\smash{{\SetFigFont{12}{14.4}{\rmdefault}{\mddefault}{\itdefault}{\color[rgb]{0,0,0}2}%
}}}}
\end{picture}%